\documentstyle[multicol,aps,epsf]{revtex}
\begin{document}
\draft
\title{Thermopower of  high-T$_c$ cuprates}
\author{Mu-Yong Choi and J. S. Kim}
\address{Department of Physics, Sungkyunkwan University, Suwon 440-746, Korea}

\date{10 September 1998}
\maketitle

\begin{abstract}
We have studied the thermopower of La$_2$CuO$_{4+z}$ and Nd$_2$CuO$_{4-y}$
which undergo an antiferromagnetic transition near room temperature
and Bi$_2$Sr$_{2-x}$La$_x$CuO$_{6+z}$ for which a broad spectrum  of doping  
is possible. The thermopower  of La$_2$CuO$_{4+z}$ and Nd$_2$CuO$_{4-y}$  is seen 
to exhibit 
an anomaly at the Neel temperature, whereas the resistivity is not.  
The extrapolated zero-temperature intercept of the  thermopower, which is known to be 
positive  
for hole-doped cuprate superconductors, is found to become negative for 
Bi$_2$Sr$_{2-x}$La$_x$CuO$_{6+z}$  above   some critical   doping-level.  The   data 
strongly suggest
that the thermopower of high-T$_c$ cuprates contains a large amount of extra
contribution in addition to the usual diffusion thermopower. 
We discuss origins of the extra contribution in the thermopower.
\end{abstract}
\draft

\begin{multicols}{2}
The nature of the normal state of the high-T$_c$ cuprate superconductors (HTSC)
has remained a key issue since their discovery. 
The normal state exhibits a variety of anomalous electronic properties.
The thermopower is one of several physical quantities which distinctly
reveal the unusual normal-state properties. The ab-plane thermopower of hole-doped
HTSC exhibits simple but unusual dependences on 
temperature and on the doping level.\cite{R1,R2,R3} At high temperatures, the 
thermopower S varies linearly in temperature T with a negative slope and a
positive zero-offset (extrapolated zero-temperature intercept).
The negative slope depends weakly on the doping level, while the zero offset 
varies from a large positive value at low doping-level to near zero in the 
overdoped region. The dependence of S on temperature and the doping level
is so systematic and universal that it can be used as a measure of the 
hole concentration in the CuO$_2$ planes for any hole-doped HTSC.\cite{R1}
Recent theoretical work\cite{R4} shows, based on the conventional Fermi-liquid 
model\cite{R5}, that the doping-level dependence of the thermopower can be explained 
by the common band-dispersion-relation of HTSC. However there does not yet
exist a plausible explanation for the unusual temperature-dependence of S, 
which is not easily reconciled with a conventional model  based on the 
usual phonon-drag contribution and/or a multi-banded electronic-structure.
The observed simplicity and universality in S seems to indicate that the 
electronic structure is simple and common to the all kinds of HTSC.  
Thermopower measurements on semiconducting and heavily-overdoped samples,
S of which has not been studied in detail yet, may provide valuable informations for
understanding the unusual T-dependence of S.

The  present   paper reports   an  investigation   of S   of  La$_2$CuO$_{4+z}$   and 
Nd$_2$CuO$_{4-y}$
which are  semiconducting and  undergo an  antiferromagnetic (AFM)  transition below 
room
temperature and Bi$_2$Sr$_{2-x}$La$_x$CuO$_{6+z}$ which enables the 
normal-state properties in the heavily-overdoped region to be studied down to below 20 
K.

The conventional solid-state reaction of stoichiometric oxides and carbonates was
adopted    in     preparing    polycrystalline    samples     of    La$_2$CuO$_{4+z}$, 
Nd$_2$CuO$_{4-y}$, 
and Bi$_2$Sr$_{2-x}$La$_x$CuO$_{6+z}$.
The X-ray diffraction analysis shows all the samples to be single phase 
within the experimental error. S was measured by employing the dc method described in 
Ref. 6. S of the polycrystalline samples represents essentially the ab-plane 
value due to the relatively high conductivity in the CuO$_2$ planes as compared 
to that along the c axis. The resistivity was measured through the 
low-frequency ac four-probe method.

\begin{figure}
\narrowtext
\centerline{\epsfxsize=3.2in
\epsffile{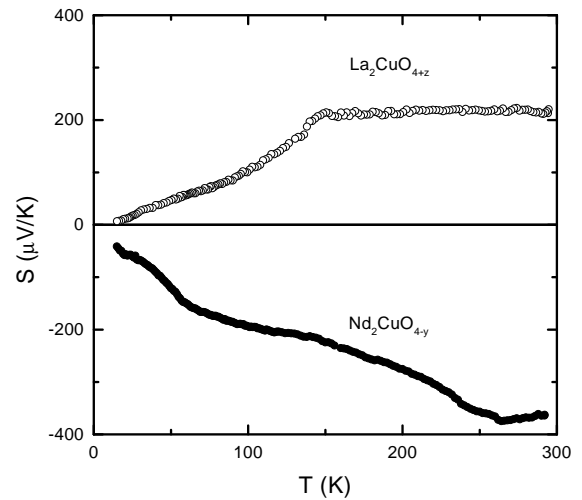}}
\vskip 0.1true cm
\caption{Temperature dependence   of the   thermopower for   La$_2$CuO$_{4+z}$ and 
Nd$_2$CuO$_{4-y}$.}
\label{fig1}
\end{figure}

\begin{figure}  
\narrowtext
\centerline{\epsfxsize=3.3in
\epsffile{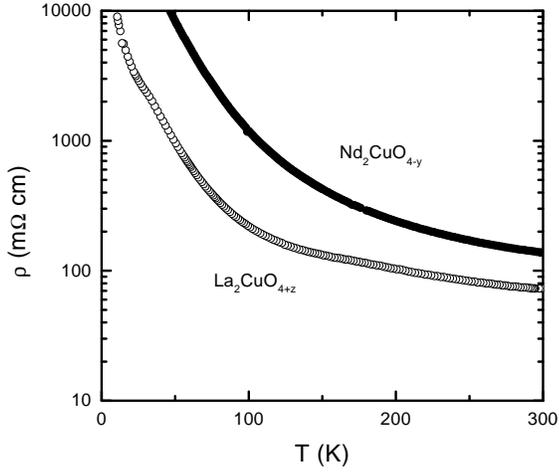}}
\vskip 0.1true cm
\caption{Temperature dependence of the resistivity $\rho$ for 
La$_2$CuO$_{4+z}$ and Nd$_2$CuO$_{4-y}$ polycrystalline samples.}
\label{fig2}
\end{figure}

Fig.\ \ref{fig1} shows the temperature dependence of S for semiconducting 
La$_2$CuO$_{4+z}$ (LCO) and Nd$_2$CuO$_{4-y}$ (NCO).
Both samples show a distinct drop of $|S|$ at $\sim${150 K} for LCO and
at $\sim$260 K for NCO. The anomaly in S is associated with 
the AFM transition\cite{R7,R8,R9}. The onset of S drop for LCO was reported to 
appear at  the same  temperature where  a sharp  peak in   the magnetic susceptibility 
appears.\cite{R10}
The onset temperature of S drop for NCO as well coincides with the
Neel temperature T$_N$ in Ref. 11.
The change of S in association with the AFM transition is as large as  $\sim50 \%$ of 
S(300 K) for  
LCO at a moderate estimate. For NCO, it is $\sim 40\%$. Despite such distinct changes 
in S associated with 
the AFM transition,  the resistivity measurements  on the  samples cut from  the same 
pellets do {\it{not}} reveal
any anomaly near T$_N$, as shown in Fig.\ \ref{fig2}.
The absence of an anomaly at T$_N$ in the resistivity has been observed in many
different experiments on LCO and NCO samples.\cite{R12,R13,R14,R15}
In the relaxation-time approximation, the diffusion contribution S$_d$ in S 
is related to the density  of states, the velocity  and the relaxation time   of conduction 
electrons, 
and so is  the resistivity  $\rho$ in  a similar  way.\cite{R5} Therefore, when   S$_d$ 
shows an anomaly 
at T$_N$,  $\rho$ is also expected to show a similar anomaly at the same temperature, 
as appeared  in transition  metals\cite{R5}. The  absence of  an anomaly  at T$_N$  in 
$\rho$ 
of our samples strongly suggests that the  observed large change in S might not  be of 
S$_d$
but of an extra contribution,  either the excitation-drag thermopower  or something else 
which is 
reduced in association of the AFM ordering.

\begin{figure}
\narrowtext
\centerline{\epsfxsize=3.3in
\epsffile{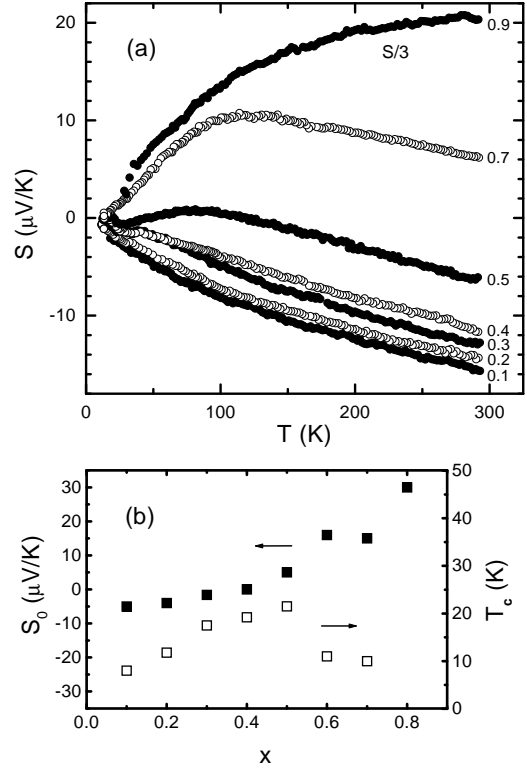}}
\vskip 0.1true cm
\caption{(a)      Temperature     dependence      of      the     thermopower      of 
Bi$_2$Sr$_{2-x}$La$_x$CuO$_{6+z}$. The 
numbers near the curves denote the lanthanum concentration $x$.
The magnitude of the thermopower of the sample with x = 0.9 is scaled down.
(b) shows the La-concentration dependence of the zero-offset thermopower S$_o$ (solid 
squares)
and the superconducting-transition temperature T$_c$ (open squares).}
\label{fig3}
\end{figure}

Fig.\ \ref{fig3} shows (a) S vs. T of Bi$_2$Sr$_{2-x}$La$_x$CuO$_{6+z}$ 
with $0.1\leq{x}\leq{0.9}$   and (b)  the  dependence on   the La-content  $x$  of the 
zero-offset 
S$_o$ and the superconducting-transition temperature T$_c$. 
The temperature and doping dependence of S for the 
samples with $x > 0.4$ are typical of HTSC; linear in T with a negative slope 
and a positive S$_o$. The zero-offset S$_o$ having a large positive value at large $x$ 
falls to zero at $x \simeq 0.4$. 
Lowering $x$ further below 0.4 (raising the hole concentration above 0.28), S$_o$  
becomes {\it{negative}}.
The hole concentration of  the sample of  $x = 0.4$  is determined from  the S(290 K) 
value and its 
correlation with the hole concentration in Ref. 1. 
Band calculations\cite{R16,R17,R18} and  photoemission experiments\cite{R19}  show that 
HTSC has 
an approximately cylindrical Fermi-surface for electrons in the CuO$_2$ planes.
An ordinary metal  with such a  simple band is  expected to have  S$_d$ linear in  T. 
Someone might 
argue that HTSC is not an ordinary metal and the positive S$_o$ might originate from
unconventional Fermi-liquid-likeness of  HTSC. The observed  development of  negative 
S$_o$ 
in the heavily-overdoped  region where  HTSC behaves  more like  an ordinary  metal, 
however, 
seems to indicate that the non-zero S$_o$ is from some 
extra contribution in S rather than unusual S$_d$.

Superconductivity has its origin at attractive electron-electron 
interactions which are mediated by some excitations interacting with electrons.
The stronger interaction between electron and the excitation generally induces the higher 
superconducting-transition temperature. Thus it would be never surprising for HTSC 
to  show  a  large  excitation-drag  thermopower,  revealing  the  presence  of  strong 
interactions 
between electron and excitations. The most ordinary excitation which  drags electrons is 
phonon.
Several authors  have tried  to explain  S of  HTSC in  terms of  phonon drag.   Early 
arguments, 
however, had some difficulty in explaining the unusual linear T-dependence of S which 
persists up to 600 K\cite{R20,R21}. Recently Trodahl\cite{R22} has  shown that inclusion 
of phonon drag and a cylindrical 
Fermi-surface can explain the unusual T-dependence within a conventional Fermi-liquid 
theory.
In the picture, the observed thermopower is a sum of  a negative S$_d$ varying 
linearly in T and a positive phonon-drag  thermopower S$_g$ varying little in  T above 100 K. 
The zero-offset 
S$_o$ is simply the saturation value of S$_g$. The observed correlation between 
S$_o$ and the doping-level is attributed to competition  between two contributions with 
opposite sign in S$_g$;
positive for the contribution from  the Umklapp processes of  electron-phonon scattering 
and 
negative for   that from   the normal  processes.  The  competition  between the   two 
contributions 
is settled by the contour of the Fermi surface which varies with the doping. 
As the hole doping is enhanced, the cylindrical hole-like Fermi-surface of
HTSC expands out and consequently the positive S$_o$ at  low doping-levels decreases 
and becomes
zero at some critical doping-level.
Extending the  argument above  the  critical level,  one might   expect that the  Fermi 
surface
ultimately turns electronlike and  S$_o$ becomes negative.  This expectation appears to 
agree qualitatively 
with our observations.

Nevertheless we note that phonon is not the only excitation  which can generate a drag 
thermopower 
and that the Trodahl's argument is not limited only for phonon drag. It can be extended 
to 
other excitations interacting with conduction  electrons, such as magnon.  It is not even 
certain for 
high-T$_c$ cuprates whether phonons interact  so vigorously with electrons.  It is well 
known for 
high-T$_c$ cuprates   that strong   electron-electron interactions   induce large   AFM 
spin-fluctuations in 
both semiconducting and  superconducting samples.  Many physicists  now believe  that 
strong interaction 
between electron   and the  spin-fluctuations  (quantum of   which is  paramagnon)  is 
responsible for the 
high-T$_c$ superconductivity.   Therefore it  could  be a   hasty conclusion  to  claim   
without extra 
evidences that phonon is the excitation.

We now examine correlation between excitation drag and the  observed S change below 
T$_N$ in the 
semiconducting  samples,  even   if similar   effects  don't  have  to   work on   both 
semiconducting and 
superconducting samples. Phonon-drag thermopower does not appear to  fit in well with 
the observation.
When phonon is the dominant excitation interacting with 
electrons, S$_g$ is not expected to change substantially in association with the 
AFM transition. It is because the electron-phonon scattering rate is
not significantly affected by the antiferromagnetic electron-spin-ordering transition.
Unlike   for   phonon,   AFM   ordering   suppresses    spin-fluctuations   and   thus 
paramagnon-drag thermopower is
reduced below T$_N$. The S charge in Fig.\ \ref{fig1} is quite similar to that observed 
in
MnTe\cite{R23}, which has been attributed to the paramagnon-drag effect. Nevertheless, 
it is not easy to 
explain why strong electron-paramagnon interaction effects would come out vividly only 
in S,
but not in $\rho$. The absence  of strong T-dependence of S above  T$_N$ associated 
with
critical  slowing   down is   another  question   to  be  answered   for admission   of 
paramagnon-drag
thermopower in the semiconducting samples.

Charge carriers in semiconducting high-T$_c$ cuprates  are known to be both  strongly 
correlated and
severely localized.   S$_d$ in  such a   system may  have,  in addition   to the  usual 
energy-transport term,
a spin-entropy term which may reach to  several hundred $\mu$V/K.\cite{R24} Liu and 
Emin\cite{R25} 
has shown that magnetic ordering reduces the spin-entropy part so effectively,
because the exchange interaction between  the carrier and the magnetic  sites limits the 
energetically 
allowable spin  configurations.  The spin-entropy  part  can be   easily reduced in   the 
presence of a large
applied magnetic field as well. Presence of a sizable spin-entropy part in S thus can be 
ascertained from
measurement of magnetothermopower. Early  measurements\cite{R26,R27} expose that S  of 
superconducting samples
is almost independent  of a magnetic  field up to  30 T. Magnetothermopower  data for 
semiconducting 
samples have not been provided yet.

In summary, we have studied S of La$_2$CuO$_{4+z}$, Nd$_2$CuO$_{4-y}$, and 
Bi$_2$Sr$_{2-x}$La$_x$CuO$_{6+z}$.       For      La$_2$CuO$_{4+z}$     and 
Nd$_2$CuO$_{4-y}$,  
S   shows   an   anomaly   at   T$_N$,   whereas   $\rho$    does   not.   For 
Bi$_2$Sr$_{2-x}$La$_x$CuO$_{6+z}$, 
the zero-offset S$_o$ is found to become negative above some critical doping-level.
The development of negative  S$_o$ in the overdoped region  looks qualitatively compatible 
with the 
excitation-drag argument for a system with a cylindrical Fermi-surface.
For the  origin of  the anomalous  change in  S below  T$_N$, paramagnon-drag  and 
spin-entropy contributions 
have been  considered.  We suggest   magnetothermopower measurement for   probe of 
spin-entropy
part in S of semiconducting samples.

We are grateful  to K.  C. Cho and  K. H.  Lim for their  assistance in  a part of  the 
experiments.

\end{multicols}

\end{document}